# Creation of a universe from thermodynamic thought experiments I

Akinbo Ojo
*Standard Science Centre*
*P.O.Box 3501, Surulere, Lagos, Nigeria*
*Email: taojo@hotmail.com*

**Abstract**

We describe a thought experiment using an isolated system of known parameters and assuming the correctness of Clausius and Boltzmann descriptions of entropy. The experiment produced an astronomical increase in the number of possible ways the system can be arranged and inescapably in achieving this, an increase in the compartment volume of the system. The result may possibly be of significance to cosmology.

**Introduction**

The correctness of Clausius description of entropy [1-4] expressed by the equation

$$\partial S = \partial E/T \qquad (1)$$

and that of Boltzmann [1-3],[5], expressed by

$$S = k \ln W \qquad (2)$$

are tested and used as a basis for experimentation, where S = entropy, E = energy, T = absolute temperature, $k$ = Boltzmann's constant (~$10^{-23}$), ln stands for natural logarithm (i.e. $\log_e$, where e = 2.71828…) and $W$ represents the different possible ways the constituents present can be arranged among the compartment units of the system. $W$ increases only if there is an increase in the number of constituents available for arrangement in different possible ways or if there is an increase in the number of compartments of the system among which that arrangement can take place.

Before proceeding to perform the experiment it may be useful to make some reference to and clarification about the thermodynamic property relation, usually written,

$$T\partial S = \partial U + P\partial V \qquad (3)$$

where U represents internal energy, P represents pressure and V the volume of the system. T and S are as previously represented above.

Pressure, P is the total force on the system boundary per unit area. In colliding with and bouncing off the system boundary and each other, the momentum of constituents in the system is reversed, momentum being a vector. Force is the rate of change of this momentum and we can find the total force and divide this by the area of the boundary\* to calculate the pressure, P present in the system. This model is based on the kinetic theory of gases and represents one of the successful models for formulating the thermodynamic laws and relations.

Newton's laws and in particular the third law are also in agreement with the model, as force (and by implication pressure) does not exist by itself if there is no reaction from without the system to oppose it. It is one of the cardinal principles of Newton's laws that for force to exist momentum must be exchanged between systems (e.g. between a gun and a bullet). For pressure to exist and do manifest work, e.g. in moving a system boundary, there must be a momentum exchange at the system's boundary possibly brought about by an opposing reaction which would come from the system's exterior. Additionally a system, where there is no displacement through which the momentum of the constituents can change cannot manifest force as change in momentum will be zero. We can therefore probably safely assume that a system which is completely isolated from any exterior system from which a reaction can come or with which it can exchange momentum and which is of smallest possible size incapable of accommodating any displacements of constituents within it to achieve changes in momentum, must on these basis have pressure within it equal to zero. For such scenarios, Eq.(3), the thermodynamic property relation reduces to Eq. (1), with the internal energy, U being the same as the E of the system. Any movement of such a system's boundary will be difficult to attribute to pressure since no reference can be made to the parameters of an exterior, the system being self-contained and oblivious of an exterior in design.

---

\*One of the important assumptions of the model in calculating pressure is that the constituents are point masses of very insignificant area when compared to the area of the system's boundary.

**The experiment**

**Aims and objectives:** To test the correctness and the applicability of some of the known and well established thermodynamic equations on the physical behavior of the universe, given that the universe may also be described as a thermodynamic system since it contains energy, entropy and an isolated volume that serves as a compartment in which constituents can be arranged in various possible ways.

**Method:** We start with a completely isolated system with a classically impervious boundary, having a single compartment and at unit absolute temperature (i.e. T = 1 kelvin).

The unit compartment can be taken to be about the Planck size and so cannot be further divided to create additional compartments among which further arrangement of constituents can be feasible.

In such a system, with only a single compartment unit available, there will be only one possible arrangement, i.e. $W = 1$, thus the initial entropy, S will be zero (see Eq.(2).

To this system we find an ingenious means of introducing a unit of energy, say one joule and evaluate the consequences of the action on the thermodynamic variables of the system.

**Results:** The following reproducible results are demonstrated as consequences for the system under experimentation.

1. The impregnation of the system with unit energy increases the entropy from zero to one, ref. Eq.(1).

2. The increase of S by one is equivalent to an increase of *W* in the system from unity to $e^{10^{23}}$, ref. Eq. (2) and taking into consideration that $k \sim 10^{-23}$.

**Discussion**

Before discussing the result of the experiment, a common misgiving is the means by which a change in energy can occur having defined the system as being classically isolated. Thought experiments sometimes allow us the indulgence to imagine seemingly impossible scenarios and subsequently try to describe what could happen based on our known physics should such seemingly impossible conjectures actually occur in reality.

We have three mechanisms to choose from to introduce energy into our system in spite of it being classically an isolated system. The first is by using the Heisenberg uncertainty principle, initially applied to the pair, position-momentum but subsequently found useful for the pair, energy-time. The principle gives us a mechanism by which a quantum of energy can realistically arise *de novo* within an isolated system without violating the laws of physics and those of energy conservation, provided such energy fluctuation does not last more than the time given by Eq.(4).

$$\Delta E \times \Delta t = \hbar \qquad (4)$$

where $\Delta E$ is the change in energy, $\Delta t$ is the duration the energy fluctuation exists and $\hbar$ is $h/2\pi$, where $h$ is Planck's constant. The use of this principle as a cosmological tool is not new. Apart from the uncertainty principle, the phenomenon of tunneling in quantum mechanics by which an entity could appear beyond a barrier from which it was classically forbidden is another possible mechanism of introducing energy through the otherwise impervious boundary of our isolated system. Thirdly, if these mechanisms are still difficult to accept, we can imagine an intelligent energetic agent opening the door to the system incredibly so slightly so it still tends towards being isolated, entering incredibly so rapidly and closing back the door incredibly so quickly that to all intents and purposes the system still practically tends towards being isolated!! In all of these what is of concern in this experiment is the observational consequence of such an eventuality taking place. Being an isolated system, the system must retain within its borders all consequences and developments from the experiment.

*W* can be described in various almost equivalent ways by different physicists, a situation that sometimes affects clarity and causes needless conflict. Being central to this discussion we may briefly note some of these. *W* may be described as the number of possible microstates consistent with the system's macroscopic thermodynamic properties; as a representation of statistical probability in terms of the position and momentum of the system's constituents, i.e. the chance of being 'here' or being 'there'; as a measure of the volume of the compartment containing the phase-space point which represents the state (this is Penrose's favorite definition [6]); as a measure of the number of different possible ways the constituents of the system can be arranged among the compartment units available in the system. Whatever means of description is preferred, the results show an astronomical change in the *W* of the system. In phase-space terms, the change in energy creates an increase in the position and momentum coordinates permissible in the system. Physically and mathematically, since there is a limit to the fragmentation of the fundamental compartment unit, such an astronomical change in *W* must occur with an increase in the number of compartment units among which the necessitated additional possible arrangements will then be made possible. An increase in the size of the system from the appearance of additional compartment units is thus imperative, with a manifest expansion or outward movement of the system boundary.

**Conclusions and inferences from the experiment**

(i) It is thermodynamically possible to obtain an astronomical increase in the value of a variable in a system from very minute changes in that system.

(ii) The volume of an isolated system can be increased from unity to a higher value and this is explicable with thermodynamic laws.

(iii) The increase in volume of the system that must occur in this particular experiment rather than being a mechanical effect from high pressure, an antigravity force, exotic particles or fields, appears mathematically and physically to be a thermodynamic effect. This being the case, a thermodynamic demand for a change in system size can be included along with other current model mechanisms for the possible explanation why an isolated system, such as our universe should increase its size from an infinitesimal or zero size to a larger size.

(iv) The experiment does not inform us whether the compartment volume increase obtained is attained instantaneously to reach an equilibrium or steady state or whether the change in compartment volume will occur gradually over time.

(v) If the result of this experiment cannot be borne out in reality despite devising a suitable mechanism that succeeds in putting energy into the system then it provides an opportunity to identify the limits and further provisos for the validity of the Clausius and Boltzmann definitions of entropy.

(vi) If the system initially composed of just one infinitesimal compartment unit later comes to be composed of multitude of such units, this may seem to confer a granular character to the body of space within the system. A notable inference from the experiment is therefore that space within the system rather than being continuous, would possess granularity as a feature.

(vii) Apart from the compelling need for additional compartment units, the increases in S and $W$ that are obtained also imply the necessity for some entity within the system to be capable of rearrangement and thus serve thermodynamically as the system's constituents. One possible candidate for such an entity capable of rearrangement in this experiment is the introduced energy, since quantum theory has demonstrated that energy is not infinitely divisible but exists in discrete units or quanta.

(viii) Taking the matter a bit further, 'rearrangement' means an identifiable change in the distribution or spatial relationship between entities. If a change in energy causes the number of compartment units in the system to vary, then the relationship of one compartment unit to others is bound to vary and in a sense we can observationally say the compartment units are capable of being rearranged. Being subject to rearrangement, the compartment units themselves may therefore also act as constituents for the system. That being the case, they may play a dual role in the thermodynamics of the system, acting as custodians of the spatial properties by conferring locus and as well conferring substantial properties on the system, by being capable of rearrangement. This is somewhat reminiscent of the 'ether' theories [7], where an ethereal substance fills the whole of space, acting as a medium in which other bodies are situated and being substantial in itself by being capable of vibrating and transmitting energy. Actually the theory survives in various modified forms in current physics, viz. the space-time background of General relativity which is capable of vibrating to produce gravitational waves and Quantum gravity proposals that space is not infinitely divisible but is granular at some infinitesimal scale, variously described as 'quantum foam', etc. Speculatively, if the compartment units can be subject to rearrangement, this implies they can be disturbed, which in turn means they can vibrate and allow energy transmission.

(ix) The experiment can be conducted for lower temperatures and correspondingly smaller amounts of energy to achieve similar results as seen from Eq.(1). For a scenario where a single quantum of energy is used for experiment with a correspondingly small temperature to achieve the same result, then since the quantum of energy used is not further divisible to give an increase in the constituents available for rearrangement, most of the necessitated increase in $W$ must come from an astronomical increase in the number of compartments in which the rearrangement of that single quantum can take place.

(x) Further interesting possibilities may exist for the speculated case where an initial absolute zero temperature is used for experimentation.

In subsequent contemplated papers we explore the possible thermodynamic origin of force, and necessarily following from this the origin of structure.

We submit this experiment for further scrutiny as it may be of importance to cosmology, whether of the big bang theory [8] or of the steady state universe theory [9], in case a similar experiment may have been performed billions of years ago.

**Acknowledgements**
I thank Prof. Animalu of the Nigerian Academy of Science and Dr. Gboyega Ojo of the University of Lagos for their encouragement. And not forgetting Stephen J. Crothers and Jeremy Dunning-Davies for new insights and criticisms through exchanged correspondence.

**References**

1. Perrot, P., *A to Z of Thermodynamics*. Oxford University Press, London, 1998

2. Fermi, E., *Thermodynamics*, Prentice-Hall, Englewood-Cliffs, 1937

3. Callen, H.B., *Thermodynamics and an Introduction to Thermostatistics*, Wiley, New York, 1985

4. Clausius, R., Uber verschiedene fur die Anwendung bequeme Formen der Hauptgleichungen der mechanischen Warmetheorie. *Annalen der Physik und Chemie*, 125:353–400, 1865.

5. Boltzmann, L., Weitere Studien uber das Warmegleichgewicht unter Gas-Molekulen. *Sitzungsbericht der Akadamie der Wissenschaften, Wien*, 66:275-370, 1872.

6. Penrose, R., *The Emperor's New Mind*, Oxford University Press, New York, 1990

7. Whittaker, E.T., *The history of the theories of aether and electricity*, Longman, London, 1910.

8. Silk, J., *The big bang, the creation and evolution of the universe*, Freeman, San Francisco, CA 1980

9. Hoyle, F., Burbidge, G and Narlikar, J., *A Different Approach to Cosmology: From a Static Universe through the Big Bang towards Reality*, Cambridge University Press, Cambridge, 2000